\documentclass[5p,authoryear,twocolumn]{elsarticle}

\usepackage{amssymb}
\usepackage{graphicx}
\usepackage{epsfig}
\usepackage{amsthm}
\usepackage{longtable}
\usepackage{lscape}
\usepackage{amsmath}

\journal{Journal of High Energy Astrophysics}

\begin{document}

\begin{frontmatter}



\title{Multimessenger Tests of Einstein's Weak Equivalence Principle and Lorentz Invariance with a High-energy Neutrino from a Flaring Blazar}



\author[a]{Jun-Jie Wei\corref{dip}}
\ead{jjwei@pmo.ac.cn}
\author[b,c]{Bin-Bin Zhang}
\ead{bbzhang@nju.edu.cn}
\author[a,d]{Lang Shao}
\author[e]{He Gao}
\author[f]{Ye Li}
\author[g]{Qian-Qing Yin}
\author[a,h]{Xue-Feng Wu}
\author[b,c]{Xiang-Yu Wang}
\author[f,i]{Bing Zhang}
\author[b,c]{Zi-Gao Dai}

\address[a]{Purple Mountain Observatory, Chinese Academy of Sciences, Nanjing 210008, China}
\address[b]{School of Astronomy and Space Science, Nanjing University, Nanjing 210093, China}
\address[c]{Key Laboratory of Modern Astronomy and Astrophysics (Nanjing University), Ministry of Education, China}
\address[d]{Department of Space Sciences and Astronomy, Hebei Normal University, Shijiazhuang 050024, China}
\address[e]{Department of Astronomy, Beijing Normal University, Beijing 100875, China}
\address[f]{Kavli Institute of Astronomy and Astrophysics, Peking University, Beijing 100871, China}
\address[g]{Key Laboratory of Particle Astrophysics, Institute of High Energy Physics, Chinese Academy of Sciences, Beijing 100049, China}
\address[h]{School of Astronomy and Space Sciences, University of Science and Technology of China, Hefei 230026, China}
\address[i]{Department of Physics and Astronomy, University of Nevada Las Vegas, Las Vegas, Nevada 89154, USA}
\cortext[dip]{Corresponding author.}

\begin{abstract}
The detection of the high-energy ($\sim290$ TeV) neutrino coincident with the flaring blazar TXS 0506+056, the first and only
$3\sigma$ neutrino-source association to date, provides new, multimessenger tests of the weak equivalence principle (WEP)
and Lorentz invariance. Assuming that the flight time difference between the TeV neutrino and gamma-ray photons from the blazar
flare is mainly caused by the gravitational potential of the Laniakea supercluster of galaxies, we show that the deviation
from the WEP for neutrinos and photons is conservatively constrained to have an accuracy of $10^{-6}-10^{-7}$, which is 3--4 orders of magnitude
better than previous results placed by MeV neutrinos from supernova 1987A. In addition, we demonstrate that
the association of the TeV neutrino with the blazar flare sets limits on the energy scales of quantum gravity for both
linear and quadratic violations of Lorentz invariance (LIV) to $E_{\rm QG, 1}>3.2\times10^{15}-3.7\times10^{16}$ GeV
and $E_{\rm QG, 2}>4.0\times10^{10}-1.4\times10^{11}$ GeV.
These improve previous limits on both linear and quadratic LIV energy scales in neutrino propagation by 5--7 orders of magnitude.
\end{abstract}


\end{frontmatter}


\def\astrobj#1{#1}
\section{Introduction}
\label{sect:intro}
On 22 September 2017, the IceCube Collaboration detected a high-energy neutrino, IceCube-170922A,
with an energy of $\sim290$ TeV \citep{2018Sci...361.1378I}. The best-fit reconstructed location is right ascension
$\rm R.A.=77.43^{+0.95}_{-0.65}$ and declination $\rm Dec.=+5.72^{+0.50}_{-0.30}$ (degrees, J2000,
90\% containment region). It was soon determined that the arrival direction of IceCube-170922A
was consistent with the location of the blazar TXS 0506+056 and coincident with a flaring state
observed since April 2017 by the Fermi-LAT \citep{2017ATel10861....1T}. The AGILE gamma-ray telescope confirmed the enhanced gamma-ray
activity at energies above 0.1 GeV from TXS 0506+056 in 10 to 23 September 2017. Follow-up
observations of the blazar led to the detection of a significant gamma-ray signal with energies up
to 400 GeV around 28 September to 4 October 2017 by the MAGIC telescope.
The significance of the temporal and spatial coincidence of the neutrino event
and the blazar flare is estimated to be at the $3\sigma$ level \citep{2018Sci...361.1378I}, which is the highest level of
confidence for cosmic neutrinos to date.
A search for further neutrinos from the direction of TXS 0506+056 in 9.5 years of IceCube data found evidence
at $3.5\sigma$ for neutrino emission in 2014--2015 \citep{2018Sci...361..147I}.
TXS 0506+056 is a blazar of BL Lacertae type and
its redshift has been recently measured to be $z=0.3365$ \citep{2018ApJ...854L..32P}.

With the physical association between the flare of TXS 0506+056 and
the high-energy neutrino, the flight time difference between the TeV neutrino and the blazar
photons can in principle be used to constrain violations of Einstein's weak equivalence principle (WEP)
and Lorentz invariance, since both their violations can lead to arrival time differences for
neutral particles of different-species or with different energies arising from the same astrophysical object.
TXS 0506+056 showed the elevated level of gamma-ray emission in the GeV band starting in April 2017,
which is prior to the IceCube-170922A alert \citep{2017ATel10861....1T}. The maximum possible
arrival-time delay between the beginning of the flare and the arrival of the neutrino is about 175 days.
On the other hand, the Fermi-LAT and AGILE observations showed that the peak of the high-energy gamma-ray flare
occurs $\sim15$ days earlier than the neutrino event.
If we assume that the neutrino event was emitted around the same time of the peak,
the time delay between the TeV neutrino and gamma-ray photons turns out to be about 15 days.

Einstein's WEP is a fundamental postulate of general relativity and other metric theories of gravity.
It states that any two different species of massless (or negligible rest mass) messenger particles
(photons, neutrinos, and gravitational waves), or any two particles of the same species but with varying
energies, if emitted simultaneously from the same astronomical source and traveling through the same
gravitational field, should reach our Earth at the same time \citep{2006LRR.....9....3W,2014LRR....17....4W}.
In the neutrino sector, the arrival time delays of MeV neutrinos and photons from supernova SN 1987A have been used to test the WEP
accuracy \citep{1988PhRvL..60..173L,1988PhRvL..60..176K} through the Shapiro (gravitational) time delay
effect \citep{1964PhRvL..13..789S}. They proved that the Shapiro delay for neutrinos is equal to that
for photons at an accuracy of 0.2-0.5\%. Assuming that the flight time difference between a PeV neutrino
and gamma-ray photons from a flare of the blazar PKS B1424-418 is mainly attributed to the gravitational potential of
supercluster, \cite{2016PhRvL.116o1101W} showed that the WEP constraint can be further improved
by two orders of magnitude. Based on the associations between five TeV neutrinos and gamma-ray photons
from gamma-ray bursts (GRBs), \cite{2016JCAP...08..031W} tightened the constraint on the deviation
from WEP to an accuracy of $\sim 10^{-11}-10^{-13}$ when adopting the gravitational potential of the
Laniakea supercluster of galaxies. Besides the neutrino-photon delays, such a test has been also applied to
the delays of photons with different energies (e.g., GRBs \citep{2015ApJ...810..121G,2016MNRAS.460.2282S,2018ApJ...860..173Y},
fast radio bursts \citep{2015PhRvL.115z1101W,2016ApJ...820L..31T}, TeV blazars \citep{2016ApJ...818L...2W},
and the Crab pulsar \citep{2016PhRvD..94j1501Y,2017ApJ...837..134Z,2018EPJC...78...86D,2018ApJ...861...66L}),
and the delays between photons and gravitational waves \citep{2016PhLB..756..265K,2016ApJ...827...75L,2016PhRvD..94b4061W,
2017ApJ...848L..13A,2017PhLB..770....8L,2017ApJ...851L..18W,2017JCAP...11..035W,2018PhRvD..97h3013S}.

Lorentz invariance is a fundamental symmetry of Einstein's relativity. However, violations of Lorentz invariance (LIV)
at the Planck energy scale $E_{\rm Pl}=\sqrt{\hbar c^{5}/G}\simeq1.22\times10^{19}$ GeV are predicted in many
quantum gravity (QG) theories attempting to unify general relativity and quantum mechanics
(see \citealt{2005LRR.....8....5M,2013LRR....16....5A}, and references therein).
As a consequence of LIV effects, the velocity of massless particles (photons or neutrinos) in a vacuum
would have an energy dependence, also known as vacuum dispersion \citep{1997IJMPA..12..607A,2007NatPh...3...87J,2008ApJ...689L...1K}.
The QG energy scale ($E_{\rm QG}$) used for representing LIV could therefore be constrained
by comparing the flight time differences of particles with different energies originating from the same source
\citep{1998Natur.393..763A,2013APh....43...50E}.
The current best limits on $E_{\rm QG}$ have been obtained from the highest energy (31 GeV) photon of
GRB 090510. The limits set are $E_{\rm QG, 1}>9.1\times10^{19}$ GeV $>(1-10)E_{\rm Pl}$
and $E_{\rm QG, 2}>1.3\times10^{11}$ GeV $>10^{-8}E_{\rm Pl}$ for linear and quadratic leading order LIV-induced
vacuum dispersion, respectively \citep{2009Natur.462..331A,2013PhRvD..87l2001V} (see also
\citealt{2011RvMP...83...11K,2013CQGra..30m3001L} and summary constraints for LIV therein).
In the neutrino sector, \cite{2008PhRvD..78c3013E} used the SN1987A MeV neutrinos to constrain the linear and quadratic
LIV energy scales, and obtained the limits of $E_{\rm QG, 1}>2.7\times10^{10}$ GeV and $E_{\rm QG, 2}>4.6\times10^{4}$ GeV.
\cite{2016PhRvL.116o1101W} analyzed possible LIV effects in neutrino propagation
from an association between a PeV neutrino and the outburst activity of blazar PKS B1424-418,
and set the limits of $E_{\rm QG, 1}>1.1\times10^{17}$ GeV and $E_{\rm QG, 2}>7.3\times10^{11}$ GeV.
Based on the associations between five TeV neutrinos and GRBs, \cite{2016JCAP...08..031W}
set the most stringent limits up to now on neutrino LIV, implying $E_{\rm QG, 1}>6.3\times10^{18}-1.5\times10^{21}$ GeV and
$E_{\rm QG, 2}>2.0\times10^{11}-4.2\times10^{12}$ GeV.

Although the tests on both the WEP and LIV have reached high precision in the neutrino sector,
most of the tests rely on the use of low-significance neutrinos correlated with photons,
which are not very reliable. Specifically, except for the MeV neutrinos from SN 1987A, the significance
of the PeV neutrino (or five TeV neutrinos) being associated with the flare of PKS B1424-418 (or GRBs) is relatively low.
The coincidences between five TeV neutrinos and GRBs only yielded a combined p-value of 0.32 \citep{2016ApJ...824..115A},
and a 5\% probability for a chance coincidence between the PeV neutrino and the PKS B1424-418 flare remains \citep{2016NatPh..12..807K}.
New high-energy neutrinos with confirmed astrophysical sources and with higher significance (e.g., this
IceCube-170922A event) are essential for further testing the WEP and LIV to a higher accuracy level.

\section{Tests of the WEP}

The motion of test particles in a gravitational field can be described by the parametrized post-Newtonian (PPN)
formalism. All metric theories of gravity satisfying the WEP predict that $\gamma_{1}=\gamma_{2}\equiv\gamma$,
where the PPN parameter $\gamma$ reflects the level of space curvature by unit rest mass and the subscripts
denotes two different particles \citep{2006LRR.....9....3W,2014LRR....17....4W}. The WEP accuracy can therefore
be characterized by constraining the differences of the $\gamma$ values for different particles.
On the basis of the Shapiro time delay effect \citep{1964PhRvL..13..789S}, the time interval required for particles
to pass through a given distance is longer by
\begin{equation}
t_{\rm gra}=-\frac{1+\gamma}{c^3}\int_{r_e}^{r_o}~U(r)dr
\end{equation}
in the presence of a gravitational potential $U(r)$,
where the integration is along the propagation path from the source $r_e$ to the observer $r_o$.
Once the WEP fails, the $\gamma$ values for different particles will no longer be the same,
resulting in the arrival-time delay of two different particles arising from the same source.
The relative Shapiro time delay is given by
\begin{equation}
\Delta t_{\rm gra}=\frac{\gamma_{\rm 1}-\gamma_{\rm 2}}{c^3}\int_{r_e}^{r_o}~U(r)dr\;,
\label{gra}
\end{equation}
where the difference of the $\gamma$ values $\Delta\gamma=\gamma_{\rm 1}-\gamma_{\rm 2}$ is deemed
as a measure of a possible deviation from the WEP.

To estimate the relative Shapiro delay with Eq.~(\ref{gra}), one needs to figure out the gravitational
potential $U(r)$. Generally speaking, $U(r)$ consists of three parts: the gravitational potentials of
our Milky Way, the intergalactic space, and the source host galaxy. For the cosmological sources,
the Shapiro delay caused by the gravitational potential of the large scale structure (e.g., nearby clusters and/or superclusters)
has been proved to be more important than the Milky Way's and the host galaxy's gravity
\citep{2016JHEAp...9...35L,2016ApJ...821L...2N,2016arXiv160104558Z}. Thus, we here consider
the gravitational potential of the Laniakea supercluster of galaxies.

Laniakea is a newly discovered supercluster of galaxies, in which our Milky Way reside \citep{2014Natur.513...71T}.
The gravitational center of Laniakea is considered as the Great Attractor \citep{1988ApJ...326...19L},
a mass concentration in the nearby universe, at a position of
$\rm R.A.=10^{h}32^{m}$, $\rm Dec.=-46^{\circ}00^{'}$.
Since the distance of TXS 0506+056 is far beyond the scale of Laniakea, the
gravitational potential of the particle paths from TXS 0506+056 to our Earth can be treated as a point mass potential
for which the Laniakea's total mass is assumed at the center of the mass.
Assuming that the observed time delay $(\Delta t_{\rm obs})$ between correlated particles
is attributed to the relative Shapiro delay, and adopting a Keplerian potential for Laniakea,
a conservative upper limit on $\Delta\gamma$ can be obtained by \citep{1988PhRvL..60..173L,2016PhRvD..94b4061W}
\begin{eqnarray}\label{eq:gammadiff}
\Delta t_{\rm obs}>\Delta t_{\rm gra}= \Delta\gamma \frac{GM_{\rm L}}{c^{3}} \times \qquad\qquad\qquad\qquad\qquad\\ \nonumber
\ln \left\{ \frac{ \left[d+\left(d^{2}-b^{2}\right)^{1/2}\right] \left[r_{L}+\left(r_{L}^{2}-b^{2}\right)^{1/2}\right] }{b^{2}} \right\}\;,
\end{eqnarray}
where $M_{\rm L}\simeq10^{17}M_{\odot}$ is the Laniakea mass \citep{2014Natur.513...71T},
$r_{L}=77$ Mpc represents the distance of the Laniakea center,
$d$ is the distance from the source (TXS 0506+056) to the Laniakea center (for a cosmic source,
$d$ can be approximated as the distance from the source to our Earth), and $b$ denotes
the impact parameter of the particle paths relative to the Laniakea center.

Since the maximum possible arrival-time delay between the beginning of the TXS 0506+056 flare
and the arrival of the neutrino is about 175 days, we first adopt 175 days as
the conservative limit of the observed time delay between the TeV neutrino and gamma-ray photons from
the blazar flare. With this time delay, the most conservative constraint on the WEP from Eq.~(\ref{eq:gammadiff}) is
$|\gamma_{\nu}-\gamma_{\gamma}|<8.5\times10^{-6}$.
Besides, much more sever limit on the WEP can be achieved ($|\gamma_{\nu}-\gamma_{\gamma}|<7.3\times10^{-7}$)
by assuming that the neutrino was emitted around the same time of the flare peak (i.e., $\Delta t_{\rm obs}\sim15$ days).
The limits on the WEP accuracy for these two assumed delays are presented in Table~\ref{table1},
which are about 3--4 orders of magnitude tighter than those of MeV neutrinos from SN 1987A.
We note that two independent works were carried out by \cite{2018arXiv180705201B} and \cite{2018arXiv180705621L},
who tested the WEP by assuming the time delay between the detected neutrino and photons is 15 days and 7 days, respectively.

\begin{table}[h]
{\normalsize
\caption{Limits on the WEP and LIV with two assumed time delays between the TeV neutrino and the blazar photons.}
\begin{tabular}{lccc}
\hline\hline
{$\Delta t_{\rm obs}$}&{$|\gamma_{\nu}-\gamma_{\gamma}|$}&{$E_{\rm QG, 1}$}&{$E_{\rm QG, 2}$}\\
(days) & & (GeV) &  (GeV)\\
\hline
175 & $8.5\times10^{-6}$ & $3.2\times10^{15}$ & $4.0\times10^{10}$ \\
15 & $7.3\times10^{-7}$ & $3.7\times10^{16}$ & $1.4\times10^{11}$ \\
\hline\hline
\end{tabular}
\label{table1}
}
\end{table}

\section{Constraints on LIV}

The LIV effect predicts an energy-dependent speed of propagation in a vacuum for neutrinos and photons.
The leading term in the modified dispersion relation for particles (with energy $E\ll E_{\rm QG}$) is
\begin{equation}
E^{2}\simeq p^{2}c^{2}+m^{2}c^{4} \pm E^{2}\left(\frac{E}{E_{\rm QG,n}}\right)^{\rm n}\;,
\label{eq:dispersion}
\end{equation}
where $m$ is the rest mass of the particle, the n-th order expansion of leading term stands for linear (n=1) or quadratic (n=2)
LIV model, and $+1$ $(-1)$ corresponds to the ``subluminal" and (``superluminal") case.
The term $m^{2}c^{4}$ among in Eq.~(\ref{eq:dispersion}) is negligible when the test particles are massless or nearly massless.
Note that the superluminal neutrinos would lose their energy rapidly
due to both vacuum pair emission and neutrino splitting \citep{2013JCAP...03..039M,2015PhRvD..91d5009S},
and excellent bounds on LIV have been made for superluminal neutrinos
\citep{2013PhRvD..87k6009B,2014PhRvD..89d3005D,2014APh....56...16S}.
Here we set the limits on subluminal neutrino LIV.
Because the speed of particles has an energy dependence, two particles with different
energies originating from the same source would arrive on Earth at different times.
The arrival-time difference due to the LIV effect is expressed as \citep{2007NatPh...3...87J}
\begin{equation}
\Delta t=\frac{1+\rm n}{2H_{0}}\frac{E_{h}^{\rm n}-E_{l}^{\rm n}}{E_{\rm QG, n}^{\rm n}}
\int_{0}^{z}\frac{(1+z')^{\rm n}dz'}{\sqrt{\Omega_{\rm m}(1+z')^{3}+\Omega_{\Lambda}}}\;,
\label{eq:tLIV}
\end{equation}
where $E_{h}$ and $E_{l}$ ($E_{h}>E_{l}$) are the energies of different particles.
Here we use the cosmological parameters obtained by the Planck observations:
$\Omega_{\rm m}=0.315$, $\Omega_{\Lambda}=0.685$, and $H_{0}=67.3$ km $\rm s^{-1}$ $\rm Mpc^{-1}$ \citep{2014A&A...571A..16P}.

Similarly, the maximum possible arrival-time delay between the beginning of the gamma-ray flare
and the arrival of the neutrino ($\sim175$ days) is firstly adopted as the upper limit of the delay
between the TeV neutrino and photons. For linear and quadratic LIV, we obtain the limits of
$E_{\rm QG, 1}>3.2\times10^{15}$ GeV and $E_{\rm QG, 2}>4.0\times10^{10}$ GeV.
With the assumption that the TeV neutrino was emitted around the same time of the flare peak,
the time delay between the neutrino and gamma-ray photons would be shorter ($\sim15$ days),
leading to much stricter limits on LIV, i.e., $E_{\rm QG, 1}>3.7\times10^{16}$ GeV and $E_{\rm QG, 2}>1.4\times10^{11}$ GeV.
The resulting constraints for these two assumed delays are summarized in Table~\ref{table1}.
Compared with the corresponding limits from MeV neutrinos of SN 1987A, our limits on $E_{\rm QG, 1}$
and $E_{\rm QG, 2}$ represent an improvement of at least 5--7 orders of magnitude.
We also note that one independent work \citep{2018arXiv180705155E} constrained linear and quadratic LIV
by assuming a difference in neutrino and photon propagation times of $\sim10$ days
(see also \citealt{2018arXiv180705621L}).

\section{Conclusions}

Very recently, a high-energy ($\sim290$ TeV) neutrino, IceCube-170922A, was detected in coincidence with
the gamma-ray emitting blazar TXS 0506+056 during an active phase, with chance coincidence being rejected
at $3\sigma$ level \citep{2018Sci...361.1378I}. This is the first time in history that confirming blazars may be
a source of cosmic neutrinos with the highest confidence level.
Based on this association between the TeV neutrino and the blazar flare,
we demonstrate that multimessenger WEP tests and neutrino LIV constraints can be carried out by using
the arrival time delay between the neutrino and the photons.
Adopting the maximum possible arrive-time difference between the neutrino and photons ($\sim175$ days),
we show that the conservative limit on the difference of the PPN $\gamma$ parameter for neutrinos and photons
is as low as $|\gamma_{\nu}-\gamma_{\gamma}|<8.5\times10^{-6}$, improving the previous WEP tests from Mev neutrinos
of SN 1987A by 3 orders of magnitude. On the other hand, we place stringent limits on linear and quadratic LIV,
namely $E_{\rm QG, 1}>3.2\times10^{15}$ GeV and $E_{\rm QG, 2}>4.0\times10^{10}$ GeV, which are
an improvement of 5--6 orders of magnitude over the previous results obtained from SN neutrinos.
If the TeV neutrino was emitted around the same time of the flare peak, the arrival-time difference between
the neutrino and photons is about 15 days. With this shorter time delay, the tests of the WEP
and Lorentz invariance can be further improved by 1 order of magnitude.

\cite{2018Sci...361..147I} investigated 9.5 years of IceCube neutrino data to search for
excess emission at the position of TXS 0506+056. An excess of high-energy neutrino events
above the expectation from the atmospheric background was found at that position during the
5-month period in 2014--2015.  Allowing for time-variable flux, this constitutes $3.5\sigma$ evidence for a neutrino
flare from the direction of TXS 0506+056, independent of and prior to the 2017 flaring
episode. This evidence supports the hypothesis presented in \cite{2018Sci...361.1378I} that the blazar TXS
0506+056 is a high-energy neutrino source. Here we show that the observed time delay of
neutrinos with different energies also provides an attractive candidate for testing
WEP and LIV. To be conservative, we adopt the duration of neutrino emission, $\sim150$ days,
as the observed time delay for neutrinos ranging in energy from about 0.1 to 20 TeV \citep{2018Sci...361..147I}.
Thus, a strong limit on the WEP from Eq.~(\ref{eq:gammadiff}) is
$|\gamma_{\nu}({\rm 20\;TeV})-\gamma_{\nu}({\rm 0.1\;TeV})|<7.3\times10^{-6}$,
and the strict limits on linear and quadratic LIV from Eq.~(\ref{eq:tLIV}) are
$E_{\rm QG, 1}>2.5\times10^{14}$ GeV and $E_{\rm QG, 2}>3.0\times10^{9}$ GeV.

\section*{Acknowledgments}
This work is partially supported by the National Basic Research Program (``973'' Program)
of China (Grant No. 2014CB845800), the National Natural Science Foundation of China
(Grant Nos. 11603076, 11673068, 11725314, U1831122, 11722324, and 11603003), the Youth Innovation Promotion
Association (2011231 and 2017366), the Key Research Program of Frontier Sciences (Grant No. QYZDB-SSW-SYS005),
the Strategic Priority Research Program ``Multi-waveband gravitational wave Universe''
(Grant No. XDB23000000) of the Chinese Academy of Sciences, and the ``333 Project'' and the Natural Science Foundation
(Grant No. BK20161096) of Jiangsu Province.

%

\begin{thebibliography}{1}
\expandafter\ifx\csname url\endcsname\relax
  \def\url#1{\texttt{#1}}\fi
\expandafter\ifx\csname urlprefix\endcsname\relax\def\urlprefix{URL }\fi
\expandafter\ifx\csname href\endcsname\relax
  \def\href#1#2{#2} \def\path#1{#1}\fi

\bibitem{2018Sci...361.1378I}
{IceCube Collaboration}, M.~G. {Aartsen}, M.~{Ackermann}, J.~{Adams}, J.~A.
  {Aguilar}, M.~{Ahlers}, M.~{Ahrens}, I.~{Al Samarai}, D.~{Altmann},
  K.~{Andeen}, et~al., {Multimessenger observations of a flaring blazar
  coincident with high-energy neutrino IceCube-170922A}, Science 361 (2018)
  eaat1378.
\newblock \href {http://arxiv.org/abs/1807.08816} {\path{arXiv:1807.08816}},
  \href {http://dx.doi.org/10.1126/science.aat1378}
  {\path{doi:10.1126/science.aat1378}}.

\bibitem{2017ATel10861....1T}
A.~J. {Tetarenko}, G.~R. {Sivakoff}, A.~E. {Kimball}, J.~C.~A. {Miller-Jones},
  {VLA Radio Observations of the blazar TXS 0506+056 associated with the
  IceCube-170922A neutrino event}, The Astronomer's Telegram 10861.

\bibitem{2018Sci...361..147I}
{IceCube Collaboration}, M.~G. {Aartsen}, M.~{Ackermann}, J.~{Adams}, J.~A.
  {Aguilar}, M.~{Ahlers}, M.~{Ahrens}, I.~A. {Samarai}, D.~{Altmann},
  K.~{Andeen}, et~al., {Neutrino emission from the direction of the blazar TXS
  0506+056 prior to the IceCube-170922A alert}, Science 361 (2018) 147--151.
\newblock \href {http://arxiv.org/abs/1807.08794} {\path{arXiv:1807.08794}},
  \href {http://dx.doi.org/10.1126/science.aat2890}
  {\path{doi:10.1126/science.aat2890}}.

\bibitem{2018ApJ...854L..32P}
S.~{Paiano}, R.~{Falomo}, A.~{Treves}, R.~{Scarpa}, {The Redshift of the BL Lac
  Object TXS 0506+056}, Astrophys. J. 854 (2018) L32.
\newblock \href {http://arxiv.org/abs/1802.01939} {\path{arXiv:1802.01939}},
  \href {http://dx.doi.org/10.3847/2041-8213/aaad5e}
  {\path{doi:10.3847/2041-8213/aaad5e}}.

\end{thebibliography}


\begin{thebibliography}{}


\bibitem[\protect\citeauthoryear{Aartsen et al.}{2016}]{2016ApJ...824..115A} Aartsen M.~G., et al., 2016, ApJ, 824, 115


\bibitem[\protect\citeauthoryear{Abbott et al.}{2017}]{2017ApJ...848L..13A} Abbott B.~P., et al., 2017, ApJ, 848, L13


\bibitem[\protect\citeauthoryear{Abdo et al.}{2009}]{2009Natur.462..331A} Abdo A.~A., et al., 2009, Natur, 462, 331


\bibitem[\protect\citeauthoryear{Amelino-Camelia et al.}{1997}]{1997IJMPA..12..607A} Amelino-Camelia G., Ellis J., Mavromatos N.~E., Nanopoulos D.~V., 1997, IJMPA, 12, 607


\bibitem[\protect\citeauthoryear{Amelino-Camelia et al.}{1998}]{1998Natur.393..763A} Amelino-Camelia G., Ellis J., Mavromatos N.~E., Nanopoulos D.~V., Sarkar S., 1998, Natur, 393, 763


\bibitem[\protect\citeauthoryear{Amelino-Camelia}{2013}]{2013LRR....16....5A} Amelino-Camelia G., 2013, LRR, 16, 5


\bibitem[\protect\citeauthoryear{Boran, Desai, \& Kahya}{2018}]{2018arXiv180705201B} Boran S., Desai S., Kahya E.~O., 2018, arXiv:1807.05201


\bibitem[\protect\citeauthoryear{Borriello et al.}{2013}]{2013PhRvD..87k6009B} Borriello E., Chakraborty S., Mirizzi A., Serpico P.~D., 2013, PhRvD, 87, 116009


\bibitem[\protect\citeauthoryear{D{\'{\i}}az, Kosteleck{\'y}, \& Mewes}{2014}]{2014PhRvD..89d3005D} D{\'{\i}}az J.~S., Kosteleck{\'y} V.~A., Mewes M., 2014, PhRvD, 89, 043005


\bibitem[\protect\citeauthoryear{Desai \& Kahya}{2018}]{2018EPJC...78...86D} Desai S., Kahya E., 2018, EPJC, 78, 86


\bibitem[\protect\citeauthoryear{Ellis et al.}{2008}]{2008PhRvD..78c3013E} Ellis J., Harries N., Meregaglia A., Rubbia A., Sakharov A.~S., 2008, PhRvD, 78, 033013


\bibitem[\protect\citeauthoryear{Ellis \& Mavromatos}{2013}]{2013APh....43...50E} Ellis J., Mavromatos N.~E., 2013, APh, 43, 50


\bibitem[\protect\citeauthoryear{Ellis et al.}{2018}]{2018arXiv180705155E} Ellis J., Mavromatos N.~E., Sakharov A.~S., Sarkisyan-Grinbaum E.~K., 2018, arXiv:1807.05155


\bibitem[\protect\citeauthoryear{Gao, Wu, \& M{\'e}sz{\'a}ros}{2015}]{2015ApJ...810..121G} Gao H., Wu X.-F., M{\'e}sz{\'a}ros P., 2015, ApJ, 810, 121


\bibitem[\protect\citeauthoryear{IceCube Collaboration et al.}{2018a}]{2018Sci...361.1378I} IceCube Collaboration, et al., 2018a, Sci, 361, eaat1378


\bibitem[\protect\citeauthoryear{IceCube Collaboration et al.}{2018b}]{2018Sci...361..147I} IceCube Collaboration, et al., 2018b, Sci, 361, 147


\bibitem[\protect\citeauthoryear{Jacob \& Piran}{2007}]{2007NatPh...3...87J} Jacob U., Piran T., 2007, NatPh, 3, 87


\bibitem[\protect\citeauthoryear{Kadler et al.}{2016}]{2016NatPh..12..807K} Kadler M., et al., 2016, NatPh, 12, 807


\bibitem[\protect\citeauthoryear{Kahya \& Desai}{2016}]{2016PhLB..756..265K} Kahya E.~O., Desai S., 2016, PhLB, 756, 265


\bibitem[\protect\citeauthoryear{Kosteleck{\'y} \& Mewes}{2008}]{2008ApJ...689L...1K} Kosteleck{\'y} V.~A., Mewes M., 2008, ApJ, 689, L1


\bibitem[\protect\citeauthoryear{Kosteleck{\'y} \& Russell}{2011}]{2011RvMP...83...11K} Kosteleck{\'y} V.~A., Russell N., 2011, RvMP, 83, 11


\bibitem[\protect\citeauthoryear{Krauss \& Tremaine}{1988}]{1988PhRvL..60..176K} Krauss L.~M., Tremaine S., 1988, PhRvL, 60, 176


\bibitem[\protect\citeauthoryear{Laha}{2018}]{2018arXiv180705621L} Laha R., 2018, arXiv:1807.05621


\bibitem[\protect\citeauthoryear{Leung et al.}{2018}]{2018ApJ...861...66L} Leung C., Hu B., Harris S., Brown A., Gallicchio J., Nguyen H., 2018, ApJ, 861, 66


\bibitem[\protect\citeauthoryear{Li et al.}{2016}]{2016ApJ...827...75L} Li X., Hu Y.-M., Fan Y.-Z., Wei D.-M., 2016, ApJ, 827, 75


\bibitem[\protect\citeauthoryear{Liberati}{2013}]{2013CQGra..30m3001L} Liberati S., 2013, CQGra, 30, 133001


\bibitem[\protect\citeauthoryear{Liu et al.}{2017}]{2017PhLB..770....8L} Liu M., Zhao Z., You X., Lu J., Xu L., 2017, PhLB, 770, 8


\bibitem[\protect\citeauthoryear{Longo}{1988}]{1988PhRvL..60..173L} Longo M.~J., 1988, PhRvL, 60, 173


\bibitem[\protect\citeauthoryear{Luo et al.}{2016}]{2016JHEAp...9...35L} Luo Z.-X., Zhang B., Wei J.-J., Wu X.-F., 2016, JHEAp, 9, 35


\bibitem[\protect\citeauthoryear{Lynden-Bell et al.}{1988}]{1988ApJ...326...19L} Lynden-Bell D., Faber S.~M., Burstein D., Davies R.~L., Dressler A., Terlevich R.~J., Wegner G., 1988, ApJ, 326, 19


\bibitem[\protect\citeauthoryear{Maccione, Liberati, \& Mattingly}{2013}]{2013JCAP...03..039M} Maccione L., Liberati S., Mattingly D.~M., 2013, JCAP, 3, 039


\bibitem[\protect\citeauthoryear{Mattingly}{2005}]{2005LRR.....8....5M} Mattingly D., 2005, LRR, 8, 5


\bibitem[\protect\citeauthoryear{Nusser}{2016}]{2016ApJ...821L...2N} Nusser A., 2016, ApJ, 821, L2


\bibitem[\protect\citeauthoryear{Paiano et al.}{2018}]{2018ApJ...854L..32P} Paiano S., Falomo R., Treves A., Scarpa R., 2018, ApJ, 854, L32


\bibitem[\protect\citeauthoryear{Planck Collaboration et al.}{2014}]{2014A&A...571A..16P} Planck Collaboration, et al., 2014, A\&A, 571, A16


\bibitem[\protect\citeauthoryear{Sang, Lin, \& Chang}{2016}]{2016MNRAS.460.2282S} Sang Y., Lin H.-N., Chang Z., 2016, MNRAS, 460, 2282


\bibitem[\protect\citeauthoryear{Shapiro}{1964}]{1964PhRvL..13..789S} Shapiro I.~I., 1964, PhRvL, 13, 789


\bibitem[\protect\citeauthoryear{Shoemaker \& Murase}{2018}]{2018PhRvD..97h3013S} Shoemaker I.~M., Murase K., 2018, PhRvD, 97, 083013


\bibitem[\protect\citeauthoryear{Stecker}{2014}]{2014APh....56...16S} Stecker F.~W., 2014, APh, 56, 16


\bibitem[\protect\citeauthoryear{Stecker et al.}{2015}]{2015PhRvD..91d5009S} Stecker F.~W., Scully S.~T., Liberati S., Mattingly D., 2015, PhRvD, 91, 045009


\bibitem[\protect\citeauthoryear{Tetarenko et al.}{2017}]{2017ATel10861....1T} Tetarenko A.~J., Sivakoff G.~R., Kimball A.~E., Miller-Jones J.~C.~A., 2017, ATel1, 10861,


\bibitem[\protect\citeauthoryear{Tingay \& Kaplan}{2016}]{2016ApJ...820L..31T} Tingay S.~J., Kaplan D.~L., 2016, ApJ, 820, L31


\bibitem[\protect\citeauthoryear{Tully et al.}{2014}]{2014Natur.513...71T} Tully R.~B., Courtois H., Hoffman Y., Pomar{\`e}de D., 2014, Natur, 513, 71


\bibitem[\protect\citeauthoryear{Vasileiou et al.}{2013}]{2013PhRvD..87l2001V} Vasileiou V., et al., 2013, PhRvD, 87, 122001


\bibitem[\protect\citeauthoryear{Wang et al.}{2017}]{2017ApJ...851L..18W} Wang H., et al., 2017, ApJ, 851, L18


\bibitem[\protect\citeauthoryear{Wang, Liu, \& Wang}{2016}]{2016PhRvL.116o1101W} Wang Z.-Y., Liu R.-Y., Wang X.-Y., 2016, PhRvL, 116, 151101


\bibitem[\protect\citeauthoryear{Wei et al.}{2015}]{2015PhRvL.115z1101W} Wei J.-J., Gao H., Wu X.-F., M{\'e}sz{\'a}ros P., 2015, PhRvL, 115, 261101


\bibitem[\protect\citeauthoryear{Wei et al.}{2016b}]{2016ApJ...818L...2W} Wei J.-J., Wang J.-S., Gao H., Wu X.-F., 2016b, ApJ, 818, L2


\bibitem[\protect\citeauthoryear{Wei et al.}{2016a}]{2016JCAP...08..031W} Wei J.-J., Wu X.-F., Gao H., M{\'e}sz{\'a}ros P., 2016a, JCAP, 8, 031


\bibitem[\protect\citeauthoryear{Wei et al.}{2017}]{2017JCAP...11..035W} Wei J.-J., et al., 2017, JCAP, 11, 035


\bibitem[\protect\citeauthoryear{Will}{2014}]{2014LRR....17....4W} Will C.~M., 2014, LRR, 17, 4


\bibitem[\protect\citeauthoryear{Will}{2006}]{2006LRR.....9....3W} Will C.~M., 2006, LRR, 9, 3


\bibitem[\protect\citeauthoryear{Wu et al.}{2016}]{2016PhRvD..94b4061W} Wu X.-F., Gao H., Wei J.-J., M{\'e}sz{\'a}ros P., Zhang B., Dai Z.-G., Zhang S.-N., Zhu Z.-H., 2016, PhRvD, 94, 024061


\bibitem[\protect\citeauthoryear{Yang \& Zhang}{2016}]{2016PhRvD..94j1501Y} Yang Y.-P., Zhang B., 2016, PhRvD, 94, 101501


\bibitem[\protect\citeauthoryear{Yu, Xi, \& Wang}{2018}]{2018ApJ...860..173Y} Yu H., Xi S.-Q., Wang F.-Y., 2018, ApJ, 860, 173


\bibitem[\protect\citeauthoryear{Zhang}{2016}]{2016arXiv160104558Z} Zhang S.-N., 2016, arXiv:1601.04558


\bibitem[\protect\citeauthoryear{Zhang \& Gong}{2017}]{2017ApJ...837..134Z} Zhang Y., Gong B., 2017, ApJ, 837, 134


\end{thebibliography}

%
\end{document}